\documentclass[aps,prd,reprint,floatfix,superscriptaddress,nofootinbib,longbibliography,a4paper]{revtex4-1}
\pdfoutput=1
\usepackage{color}
\usepackage{graphicx}
\usepackage{textcomp}
\usepackage{tabularx}
\usepackage{subfigure}
\usepackage{feynmp}
\usepackage{amsmath,amssymb}
\usepackage{enumerate}
\usepackage{slashed}
\usepackage[normalem]{ulem}
\usepackage{hhline}
\usepackage{hyperref}
\hypersetup{colorlinks=true,
    citecolor=blue,
	linkcolor=blue,
	filecolor=magenta,      
	urlcolor=blue}
\newcommand{\mathsym}[1]{{}}

\newcommand{\beqa}{\begin{eqnarray}}
\newcommand{\eeqa}{\end{eqnarray}}
\newcommand{\be}{\begin{equation}}
\newcommand{\ee}{\end{equation}}
\newcommand{\ba}{\begin{array}} 
\newcommand{\ea}{\end{array}}

\begin{document} 
\title{ Dominant One-Loop Seesaw Contribution Induced by Non-Invertible Fusion Algebra}
\bigskip
\author{Monal Kashav}
\email{monalkashav@gmail.com}
\affiliation{Theoretical Physics Division, Physical Research Laboratory, Navarangpura, Ahmedabad-380009, India}

\
\begin{abstract}
\noindent The topological classification of the one-loop Weinberg operator at dimension-5 enables a systematic categorization of radiative neutrino mass models. Among these, the category consisting loop-extended seesaw frameworks is theoretically appealing but conventional discrete or continuous symmetries (\emph{e.g.}, $U(1)$ or $\mathbb{Z}_M$) cannot genuinely forbid the corresponding tree-level contributions, making loop dominance difficult to realize. We show that \textit{non-invertible selection rules} (NISRs) naturally enforce the absence of tree-level terms while ensuring a dominant one-loop contribution. Intriguingly, the same non-invertible structure also stabilizes the dark matter candidate, providing a unified radiative origin of neutrino mass and dark sector stability. In particular, we focus on the T4-2-$i$ topology which embodies a type-II one-loop seesaw and demonstrate its natural realization from \(\mathbb{Z}_{7}\) Tambara--Yamagami (TY) fusion algebra.
\end{abstract}

\maketitle

\section{\label{sec:1}Introduction}

\noindent The dimension-five Weinberg operator $(LLHH)/\Lambda$ provides the simplest origin of lepton number violation $(\Delta L=2)$ and explains the smallness of neutrino masses~\cite{Weinberg:1979sa}. Its ultraviolet completions lead to the canonical Type-I~\cite{Minkowski:1977sc, Mohapatra:1979ia}, Type-II~\cite{Magg:1980ut, Schechter:1980gr, Wetterich:1981bx}, and Type-III~\cite{Foot:1988aq} seesaw frameworks, where heavy singlet or triplet mediators generate suppressed neutrino masses. However, these mechanisms typically invoke super-heavy states and fine-tuned Yukawa structures, placing them beyond current collider reach. Radiative realizations offer a more natural low-scale alternative, where neutrino masses arise through loop-induced operators often linked to dark matter stability. The loop suppression allows mediators near the electroweak scale, leading to richer phenomenology.

\noindent A systematic understanding of radiative neutrino mass models arises from the topological classification of the dimension-5 Weinberg operator at one and two loops~\cite{Bonnet:2012kz, AristizabalSierra:2014wal}. At one loop, six distinct topologies (T$_i$, $i=1,\ldots,6$) with four external legs exhaust all possible diagrammatic realizations. These topologies can be organized into three classes:  
(i) \textit{divergent one-loop extensions of seesaw} (e.g. T4-1-$i$, T4-2-$ii$, T$5$, T$6$),  
(ii) \textit{finite one-loop diagrams with naturally leading contributions} (T1-$i$, T1-$ii$, T1-$iii$, T$3$), and  
(iii) \textit{finite one-loop extensions of tree-level seesaw} (T4-1-$ii$, T4-2-$i$, T4-3-$i$).  

\noindent The topologies grouped under category~(iii) correspond to finite one-loop extensions of the canonical seesaw frameworks. These can be summarized as follows:
\begin{itemize}
 \item \textbf{Type-I/III one-loop seesaw:} realized through the topology T4-3-$i$,
    \item \textbf{Type-II one-loop seesaw:} realized through the topological structures T4-1-$ii$ and T4-2-$i$.
   
\end{itemize}
A natural question then arises: can a simple symmetry, such as a discrete $\mathbb{Z}_N$ or continuous $U(1)$, forbid the unwanted tree-level contribution while preserving the desired one-loop term? For the T$_1$ and T$_3$ topologies, discrete or $U(1)$ symmetries can effectively suppress tree-level contributions, rendering the one-loop term dominant. However, symmetry-based analyses reveal that for the T$_4$ class, the tree-level term contributing to neutrino mass invariably reappears, regardless of the imposed $\mathbb{Z}_N$ or $U(1)$ charge assignments \cite{Bonnet:2012kz}. Consequently, these conventional (invertible) symmetries fail to realize a genuinely dominant one-loop contribution, highlighting the need for a more general  symmetry structure to achieve radiative neutrino mass generation in its simplest form. Few attempts have been made to deal with it using flavor symmetries such as $D_4$ in addition to combinations of cyclic $\mathbb{Z}_M$ groups with enlarged field content~\cite{Loualidi:2020jlj} and with modular $A_4$ symmetry having freedom in choice of modular weights~\cite{Kashav:2022kpk}. This limitation makes it challenging to realize genuine loop dominance without fine-tuning or ad-hoc field extensions. Nonetheless, these symmetries remain \textit{invertible}, and the corresponding selection rules are not robust enough to prevent the reappearance of the tree-level Weinberg operator once lepton number is violated. This motivates exploring a fundamentally different class of symmetry principles.  

\noindent More recently, an expanded theoretical landscape has emerged following the recognition that non-invertible symmetries -- defined by the presence of topological defects whose fusion rules do not form a group but rather a generalized fusion algebra -- can furnish selection rules far more restrictive than those obtained from ordinary discrete symmetries. This development has led to a substantial body of formal work on the construction, classification and symmetry-topological field theory (TFT) realisation of non-invertible defects~\cite{Shao:2023gho,Bhardwaj:2022kot,Kaidi:2022cpf}, accompanied by a growing phenomenological implications for fermion mass textures, Yukawa hierarchies, and radiative neutrino-mass generation ~\cite{Kobayashi:2024yqq,Kobayashi:2024cvp,Kobayashi:2025znw,Suzuki:2025oov,Kobayashi:2025ldi,Liang:2025dkm,Kobayashi:2025cwx,Kobayashi:2025lar,Nomura:2025sod,Chen:2025awz,Nomura:2025yoa,Dong:2025jra,Okada:2025kfm,Kobayashi:2025thd,Suzuki:2025bxg,Jangid:2025krp,Kobayashi:2025rpx,Jiang:2025psz,Nomura:2025tvz,Suzuki:2025kxz}. Motivated by these advances, we propose that \textit{non-invertible symmetries}, characterized by their non-group-like fusion algebras, can provide a structurally natural mechanism for resolving this problem. In this work, we propose that \textit{non-invertible symmetries}, characterized by fusion algebras, can elegantly resolve this issue.In particular, non-invertible selection rules (NISRs) arising from derived from \(\mathbb{Z}_7\) Tambara-Yamagami (TY) fusion algebra naturally forbid the tree-level operators while allowing the one-loop realization of neutrino masses. As a concrete realization, we focus on the T4-2-$i$ topology, corresponding to a Type-II one-loop seesaw framework implemented under TY selection rules. This setup not only yields a dominant radiative contribution to neutrino masses but also naturally accommodates viable dark matter candidates whose stability is intrinsically ensured by the underlying non-invertible symmetry structure.

\section{\label{sec:2} One-Loop Topologies under Conventional and Non-Invertible Symmetries}

\noindent In this section, the implementation of one-loop neutrino mass generation is examined under both conventional invertible (group-based) and non-invertible symmetry structures. The discussion first clarifies the algebraic distinctions between invertible fusion rules and their non-invertible generalizations, followed by an assessment of the implications of standard symmetries such as \(\mathbb{Z}_N\) or \(U(1)\) for the radiative topology T4-2-i. A corresponding analysis is presented for the non-invertible case, highlighting how the associated fusion relations qualitatively modify the operator structure and selection rules within the same framework.

\subsection{\label{sec:2A}Fusion Algebras: Invertible vs.\ Non-Invertible}

\noindent
In general, discrete or continuous symmetries such as $\mathbb{Z}_N$ or $U(1)$ are described by invertible group structures, where each element possesses an inverse and all possible charge assignments are additive modulo $N$. The generator $g$ of $\mathbb{Z}_N$ satisfies
\[
g^N= [e], \qquad g^{i} g^{j} = g^{i+j},
\]
and all elements commute, reflecting an abelian and hence fully invertible structure. Physical fields $\phi$ carrying charge $k$ under $\mathbb{Z}_N$ transform as
\[
\phi \; \longrightarrow \; e^{2\pi i k/N} \, \phi,
\]
leading to the familiar charge-conservation condition for allowed interactions,
\[
g^{k_1} g^{k_2} g^{k_3} = e \quad \Rightarrow \quad k_1 + k_2 + k_3 \equiv 0 \ (\mathrm{mod}\ N),
\]
which defines the standard \emph{selection rule}.
\noindent
A natural extension beyond conventional $\mathbb{Z}_N$ or $U(1)$ symmetries arises when one considers operations acting not only on the fields but also on the symmetry group itself. Such \emph{automorphisms}, e.g., $g \mapsto g^{-1}$, reflect meta-symmetry transformations that preserve group structure yet act nontrivially on its elements. Gauging this $\mathbb{Z}_2$ automorphism effectively identifies $[g^k]$ with $[g^{-k}]$, reducing the independent sectors and promoting the resulting symmetry algebra to a \emph{non-invertible fusion category} as discussed in \cite{Kobayashi:2024yqq}.

\noindent Here, we focus on Tambara - Yamagami (TY) categories (for details see Appendix~\ref{app:A}) which provide the minimal non-invertible extension of a $\mathbb{Z}_N$ symmetry by adjoining a single new object $X$ whose self-fusion produces a direct sum over all group elements \cite{Tambara:1998vmj}. This construction preserves the invertible $\mathbb{Z}_N$ subsector while introducing non-invertible behaviour exclusively through $X$. The characteristic fusion rule $X\otimes X=\bigoplus_{g\in\mathbb{Z}_N} g$ leads to selection rules qualitatively different from any group symmetry: bilinears involving $X$ may be allowed through individual group channels, whereas trilinears are often suppressed. In this way, TY categories generalize $\mathbb{Z}_N$ symmetries in a controlled and physically meaningful manner.


\subsection{\label{sec:2B}Discrete Invertible Symmetry Constraints on Loop Dynamics}
\begin{figure}
    \centering
    \includegraphics[width=0.7\linewidth]{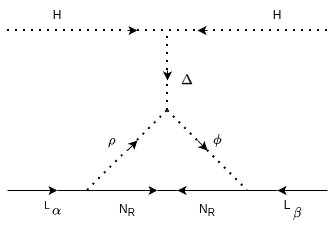}
    \caption{Schematic of general T4-2-i topological diagram corresponding to one--loop Type-II seesaw.}
    \label{fig:1}
\end{figure}
\noindent For discrete symmetries such as $\mathbb{Z}_N$, it has been shown that how charge assignments control the one-loop topology T4-2-$i$ \cite{Bonnet:2012kz}. As shown schematically in Fig.~\ref{fig:1}, all interaction vertices must be invariant under the imposed symmetry. Denoting the charge of each field by $q_j$, the invariance of the trilinear vertex $\mu_H H \Delta H$ gives the condition
\[
2q_H + q_\Delta = 0.
\]
Similarly, for $\mu_\Delta \Delta \rho \phi^{\dagger}$ and the Yukawa couplings $y,\,y'$, one obtains
\[
q_\Delta - q_\rho + q_\phi = 0, \qquad q_L + q_N - q_\rho = 0, \qquad q_L - q_N + q_\phi = 0.
\]
Combining these relations leads to $2q_L + q_\Delta = 0$, which means that the tree-level seesaw contribution $L_{\alpha}\Delta L_{\beta}$ is automatically allowed whenever the trilinear term $ H \Delta H$ is permitted. In invertible symmetries, where charges simply combine linearly modulo $N$, this overlap between tree-level and loop-level terms cannot be avoided. Therefore, isolating the radiative one-loop contribution within the T4-2-$i$ topology requires a framework that extends beyond additive conservation. 

\begin{table*}[t]
\centering
\caption{Field content and discrete symmetry assignments. Fields are in the top row, with corresponding discrete representations in the bottom row.}
\label{tab:fields}
\begin{tabular*}{\textwidth}{@{\extracolsep{\fill}}ccccccccccc}
\hline\hline
$\bar{L}_e$ & $\bar{L}_\mu$ & $\bar{L}_\tau$ & $l_e$ & $l_\mu$ & $l_\tau$ & $H$ & $\Delta$ & $\rho$ & $\phi$ & $N_R$ \\
\hline
$g$ & $g^2$ & $g^3$ & $g^6$ & $g^5$ & $g^4$ & $1$ & $1$ & $X$ & $X$ & $X$ \\
\hline\hline
\end{tabular*}
\end{table*}

\subsection{\label{sec:2C}Non-Invertible Fusion Effects on T4-2-i Realizations}
\noindent In contrast to ordinary additive symmetries, non-invertible symmetry relations (NISRs) allow for fusion rules among symmetry labels that are intrinsically non-additive. A useful and minimal framework for realizing such behavior is provided by the Tambara--Yamagami fusion category based on the group \( \mathbb{Z}_7 \). In this construction, described in Appendix~\ref{app:A}, the fusion rules are determined not simply by group multiplication but by the richer structure
\[
X \otimes X = 1 + g + g^2 + g^3 + g^4 + g^5 + g^6,\quad g^{i} \otimes X = X \otimes g^{i} = X,
\]
where \(g^{i}\) are the invertible elements of \( \mathbb{Z}_7 \) and \(X\) is the unique non-invertible object characterizing the category.

\medskip

\noindent
The field content of the model and their discrete symmetry assignments are summarized in Table~\ref{tab:fields}. Here, $L_e$, $L_\mu$, and $L_\tau$ denote the three generations of left-handed leptons, while $l_e$, $l_\mu$, and $l_\tau$ correspond to the right-handed charged leptons. The scalar fields $H$, $\phi$, and loop mediators $\Delta$ and $\rho$ are also listed along with the one right-handed neutrino $N_R$. The discrete assignments are chosen such that the Standard Model fields remain invertible under the symmetry, whereas fields in the loop $\phi$, $\rho$ and $N_R$ are non-invertible, which allows them to mediate loop-level interactions without generating tree-level contributions that are forbidden by the fusion rules.

\noindent
The full Lagrangian can therefore be expressed as
\begin{equation}
    \mathcal{L} = \mathcal{L}_{\rm SM} + \mathcal{L}_{\rm Yuk} + \mathcal{L}_{\rm Scalar},
\end{equation}
The Yukawa Lagrangian relevant for the masses of charged leptons and neutrinos is given by
\begin{align}
\mathcal{L}_{\rm Yuk} &= Y_e \, \bar{L}_e \, l_e \, H + Y_\mu \, \bar{L}_\mu \, l_\mu \, H + Y_\tau \, \bar{L}_\tau \, l_\tau \, H \nonumber \\
&\quad + r_1 \,\bar{L}_e \, N_R \, \rho + r_2 \, \bar{L}_\mu \, N_R \, \rho + r_3 \, \bar{L}_\tau \, N_R \, \rho  \nonumber \\
& \quad + p_1 \, \bar{L}_e \, N_R \, \phi + p_2 \, \bar{L}_\mu \, N_R \, \phi + p_3 \, \bar{L}_\tau \, N_R \, \phi, \nonumber \\
& \quad + M \bar{N}_R^c N_R +  {\rm h.c.}  \nonumber \\
\end{align}

\noindent Loop vertex involving the non-invertible fields $\rho$ and $\phi$ are allowed, such as $\Delta \rho \phi$, since $1 \otimes X \otimes X \supset 1$ contains the identity element in the fusion algebra. This operator is crucial for generating higher-order contributions to the neutrino mass matrix and for connecting the non-invertible sector with the Standard Model fields. At the same time, tree-level operators such as $L_\alpha L_\beta \Delta$ or $\bar{L}_\alpha \tilde{H} N_{R}$ are forbidden, since the fusion of their representations does not contain the identity. This separation between allowed and forbidden terms highlights the selective nature of non-invertible symmetry constraints. Owing to the Table~\ref{tab:fields}, the scalar vertices allowed are
\begin{equation}
    \mathcal{L}_{\rm Scalar} = \lambda_1 H \Delta H + \lambda_2 \, \Delta \, \rho \, \phi +  {\rm h.c.} 
\end{equation}
\noindent
\noindent After spontaneous symmetry breaking, the discrete fusion rules constrain the Yukawa sector such that only the diagonal terms $\bar{L}_\alpha l_\alpha H$ $(\alpha = e,\mu,\tau)$ remain invariant, as their fusion products contain the identity. Off-diagonal structures $\bar{L}_\alpha l_\beta H$ $(\alpha \neq \beta)$ are excluded since their fusion channels lack the identity element. Consequently, the charged-lepton mass matrix is forced to be diagonal,
\[
M_\ell = v_H\, \mathrm{diag}(Y_e,\, Y_\mu,\, Y_\tau),
\]

\noindent
where, $v_H$ is the vacuum expectation value of Higgs. Similarly, the neutrino Dirac couplings are constrained. Couplings of the left-handed leptons to $N_R$ mediated by $\rho$ and $\phi$ are allowed because the fusion of $L_\alpha \otimes N_R \otimes \rho$ ($\textrm{or}$ $\phi$)  contains the identity.  After SSB, neutrino Dirac couplings are given by

\begin{equation}
Y_\nu^{\rho} \equiv r=
\begin{pmatrix}
r_1 \\
r_2 \\
r_3
\end{pmatrix}, \qquad
Y_\nu^{\phi} \equiv p=
\begin{pmatrix}
p_1 \\
p_2 \\
p_3
\end{pmatrix}.
\label{eq:Ynus}
\end{equation}
The Yukawa structures $Y_\nu^{\rho}$ and $Y_\nu^{\phi}$ capture the flavour-dependent interactions 
of the fermion $N_{R}$ with the two scalar mediators $\rho$ and $\phi$. The loop function contains the full mass dependence arising from the 
propagators of the mediators $(\rho,\phi)$ and the internal Majorana fermion $N_R$. With these ingredients, 
the one--loop contribution to neutrino mass is \cite{Bonnet:2012kz}
\begin{equation}
M_\nu \;=\; - \sum_{k} 
\frac{\lambda_{1}{v_H}^{2}}{M_{\Delta}^{2}\,\lambda_{2}}\,
\bigl[\, Y_{\rho}(M_{k})\, Y_{\phi}^{T} \,\bigr]\,
I\!\left(M_{\rho}^{2},\, M_{\phi}^{2},\, M^{2}\right),
\end{equation}
where, $M$, $M_\rho$, $M_\phi$ and $M_\Delta$ are masses of $N_R$, $\rho$, $\phi$ and $\Delta$ fields. 
\noindent A distinctive feature of this construction is that it generates atleast two non-degenerate neutrino masses using only a single generation of right-handed neutrino, in contrast to conventional radiative models that require multiple sterile states. This is a direct consequence of the asymmetric structure
\begin{equation}
M_\nu \propto 
Y_\nu^{\rho}\,(Y_\nu^{\phi})^{T}
+ 
Y_\nu^{\phi}\,(Y_\nu^{\rho})^{T},
\end{equation}
whose mixed bilinear form guarantees $\det(M_\nu)\neq 0$.  
Moreover, with only one right-handed neutrino, the loop function decouples from the flavour structure and factors out as a universal coefficient, 
\(I(M_\rho^{2},\,M_\phi^{2},\,M^{2})\),  
leaving the entire mass matrix determined solely by the Yukawa bilinears which explicitly take the form
\begin{equation}\label{mnu}
M_\nu 	\propto
\begin{pmatrix}
2 r_1 p_1 & r_1 p_2 + r_2 p_1 & r_1 p_3 + r_3 p_1 \\
r_1 p_2 + r_2 p_1 & 2 r_2 p_2 & r_2 p_3 + r_3 p_2 \\
r_1 p_3 + r_3 p_1 & r_2 p_3 + r_3 p_2 & 2 r_3 p_3
\end{pmatrix}.
\end{equation}

\noindent The neutrino mass matrix in Eqs.~(\ref{mnu}) is a sum of two rank--one
contributions and therefore has rank at most two i.e. one neutrino is massless (\(m_1\) or \(m_3=0\)).  Introducing the three
bilinears
\(
a = r_1^2 + r_2^2 + r_3^2,\,
b = r_1 p_1 + r_2 p_2 + r_3 p_3,\,
c = p_1^2 + p_2^2 + p_3^2,
\)
the two nonzero eigenvalues of $M_\nu$, for the case \(m_1=0\) are
\begin{equation}
m_{2,3} = b \,\pm\, \sqrt{\,a c\,},
\label{eq:evals}
\end{equation}
reflecting the
fact that $M_\nu$ acts nontrivially only in the two - dimensional subspace
spanned by the vectors $r$ and $p$. Consequently, mass-squared differences are given by Hence
\begin{equation}
\Delta m_{21}^2=(b-\sqrt{ac})^2,\qquad
\Delta m_{32}^2=4b\sqrt{ac}.
\label{eq:dmsq_compact}
\end{equation}

\noindent Since the nonzero eigenvectors lie in ${\rm span}\{r,p\}$, an eigenvector
of $M_\nu$ with eigenvalue $\lambda$ has the form
$v = \alpha' r + \beta' p$.  
Solving the eigenvalue equation gives the coefficient ratio
\begin{equation}
\frac{\beta'}{\alpha'}
= \frac{\lambda - b}{c}
= \pm\,\sqrt{\frac{a}{c}},
\qquad
\lambda = m_{2,3},
\label{eq:ratio}
\end{equation}
so the corresponding eigenvectors can be written as
\begin{equation}
U_{\alpha 2,\alpha 3}
= r \,\pm\,\sqrt{\frac{a}{c}}\,\, p
= 
\begin{pmatrix}
r_1 \pm \sqrt{\frac{a}{c}}\, p_1 \\
r_2 \pm \sqrt{\frac{a}{c}}\, p_2 \\
r_3 \pm \sqrt{\frac{a}{c}}\, p_3
\end{pmatrix},
\label{eq:vpm}
\end{equation}
such that $ s\equiv\sqrt{\frac{a}{c}}>0$. The massless eigenvector is orthogonal to both $r$ and $p$ and is thus
proportional to their cross product.  In components this reads
\begin{equation}
U_{\alpha 1} \;=\; r \times p
=
\begin{pmatrix}
r_2 p_3 - r_3 p_2 \\
r_3 p_1 - r_1 p_3 \\
r_1 p_2 - r_2 p_1
\end{pmatrix},
\quad
M_\nu\, U_{\alpha 1} = 0.
\label{eq:v0}
\end{equation}

\noindent After normalization, the three eigenvectors form the diagonalizing matrix
\begin{equation}
U_\nu =
\begin{pmatrix}
\displaystyle \frac{U_{\alpha 1}}{\|U_{\alpha 1}\|} &
\displaystyle \frac{U_{\alpha 2}}{\|U_{\alpha 2}\|} &
\displaystyle \frac{U_{\alpha 3}}{\|U_{\alpha 3}\|}
\end{pmatrix},
\label{eq:Ufinal}
\end{equation}
such that 
\begin{equation}
    U_\nu^{T} M_\nu\, U_\nu
= \mathrm{diag}(0,m_2,m_3) \quad\textrm{or} \quad\mathrm{diag}(m_1,m_2,0),
\end{equation}
depending on normal or inverted hierarchy of neutrinos with squared norms of the eigenvectors corresponding to eigenvalue $m_2,m_3$ as
\(\|r\pm s p\|^2=2\bigl(a\pm s b\bigr)\). Using the PDG parameterization, the mixing angles are given by
\begin{align}
\sin^2\theta_{13} &= |U_{e3}|^2
= \frac{(r_1 + s p_1)^2}{2(a + s b)}, \label{eq:s13_compact}\\[4pt]
\sin^2\theta_{23} &= \frac{|U_{\mu3}|^2}{1-|U_{e3}|^2}
= \frac{(r_2 + s p_2)^2}{2(a + s b) - (r_1 + s p_1)^2}, \label{eq:s23_compact}\\[4pt]
\sin^2\theta_{12} &= \frac{|U_{e2}|^2}{1-|U_{e3}|^2}
= \frac{\dfrac{(r_1 - s p_1)^2}{2(a - s b)}}
{1 - \dfrac{(r_1 + s p_1)^2}{2(a + s b)}}\,. \label{eq:s12_compact}
\end{align}

\noindent This construction clearly distinguishes the roles of invertible Standard Model fields and non-invertible loop mediators, ensuring that the symmetry constraints are respected while generating a viable structure for charged-lepton and neutrino masses.
\noindent The analytical results presented are derived
under the simplifying assumption that the parameters $r_i$ and $p_i$
are real. In general, however, these quantities are complex, although
not all complex phases are independent. By performing the phase
redefinition of leptonic fields (for details see Appendix \ref{app:B}), the unphysical phases can be removed. Further, the neutrino mass matrix
$M_\nu$ is diagonalized and the corresponding low-energy observables
are computed. The resulting values are summarized in
Table~\ref{tab:num_results}.

\begin{table}[t]
\centering
\caption{Theory parameters and resulting neutrino observables for normal hierarchy (NH).}
\label{tab:num_results}
\renewcommand{\arraystretch}{1.15}
\begin{tabular}{c c | c c}
\hline\hline
\multicolumn{2}{c|}{\textbf{Input Parameters}} 
& \multicolumn{2}{c}{\textbf{Observables}} \\
\hline
$r_1$ & $0.219$ 
& $\Delta m^2_{21}/\Delta m^2_{31}$ & $0.0280$ \\

$r_2$ & $0.891$ 
& $\sin^2\theta_{12}$ & $0.307$ \\

$r_3$ & $0.900$ 
& $\sin^2\theta_{13}$ & $0.0220$ \\

$p_1$ & $0.368 + 0.207\,i$  
& $\sin^2\theta_{23}$ & $0.561$ \\

$p_2$ & $-0.442 + 0.897\,i$ 
& $\delta_{\rm CP}$ & $1.15$ \\

$p_3$ & $-0.823 - 0.173\,i$ 
& $\alpha$ & $1.09$ \\

 &  
& $\beta$ & $1.65$ \\

 &  
& $m_1$ [eV] & $0$ \\

 &  
& $m_2$ [eV] & $0.00861$ \\

 &  
& $m_3$ [eV] & $0.0514$ \\

 &  
& $m_{\beta\beta}$ [eV] & $0.00265$ \\

\hline\hline
\end{tabular}
\end{table}

This demonstrates that the chosen complex parameters $\{r_i,p_i\}$ reproduce all neutrino observables, including masses, mixing angles, and
CP-violating phases. The sum of neutrino masses
remains well below the current cosmological upper bound, and is therefore fully
consistent with the latest limit $\sum m_\nu < 0.064~\mathrm{eV}$ (95\% C.L.)
reported by the DESI collaboration \cite{DESI:2025zgx}.

\section{Dark Matter Stability from the Non-Invertible Element\texorpdfstring{$X$}{X}}
\noindent
The non-invertible fusion structure plays a decisive role in stabilizing the dark sector. 
In this model, the loop fields $\rho$, $\phi$, and the right-handed neutrino $N_R$ all carry the non-invertible label $X$, while the SM fields reside in invertible $\mathbb{Z}_7$ representations.  
The Tambara--Yamagami rules imply that $X$ cannot fuse with an SM representation to yield the identity.

\medskip
\noindent
As a result, interactions of the form 
\[
\bar{L}_\alpha\,l_\beta\,\rho, \quad 
\bar{L}_\alpha\,l_\beta \,\phi
\]
are automatically forbidden: no admissible fusion path connects an invertible SM object to a single $X$.  
Therefore, $\rho$ and $\phi$ never generate charged-lepton contributions, keeping all their effects confined to the neutrino loop.  
This selective coupling is an intrinsic outcome of non-invertibility, since $X$ has no fusion channel that reduces to purely in SM representations.

\medskip
\noindent  
Depending on the mass ordering, the model thus accommodates either fermionic ($N_R$) or scalar ($\rho$ or $\phi$) dark matter.  
In summary, $X$ simultaneously forbids unwanted SM couplings and protects the dark sector, providing stability without introducing an extra $\mathbb{Z}_2$ symmetry.

\section{Summary AND Discussions}
\noindent Here, we have shown, how the inclusion of a single non-invertible object enriches the symmetry structure, yielding selection rules that are different from the framework of purely invertible (group-like) symmetries. This structure highlights two central features. First, all Standard Model fields and seesaw field (\(\Delta\)) are assigned invertible, \(\mathbb{Z}_7\)-like charges. As a result, operators involving only SM  and seesaw fields obey conventional additive selection criteria, analogous to those present in ordinary discrete symmetries. Second, the non-invertible charge \(X\) is carried exclusively by the loop-mediating fields \(\rho\), \(\phi\), and \(N_R\). This assignment leads to a qualitative distinction in operator admissibility: interactions involving pairs of non-invertible fields fuse to a direct sum that includes the identity object, enabling loop-induced invariants that would be otherwise inaccessible under a purely invertible symmetry. Thus, a minimal non-invertible extension of \(\mathbb{Z}_7\) suffices to reshape the operator spectrum, suppressing tree-level contributions while maintaining the radiative pathway—an outcome that cannot be achieved within the framework of invertible symmetries alone. Also, Non-invertible element $X$ stabilizes the dark matter candidates on careful identification.

\begin{acknowledgments}
\noindent This work is supported by the Department of Space (DOS), Government of India. MK would like to acknowledge Physical Research Laboratory, Ahmedabad for the Post-Doctoral fellowship.
\end{acknowledgments}

\appendix
\section{\label{app:A}Non-Invertible TY Fusion Algebra}
\noindent Tambara--Yamagami (TY) categories were introduced by Tambara and Yamagami in classification of a distinguished class of spherical fusion categories built from a finite Abelian group augmented by a single non-invertible object. Their construction emerged from broader efforts to understand semisimple tensor categories beyond group-theoretic frameworks, influenced by progress in subfactor theory, operator algebras, and the early classification of low-rank fusion categories. TY categories represent one of the earliest explicit realizations of non-invertible fusion, characterized by a non-invertible object of quantum dimension $\sqrt{|G|}$ whose fusion closes only through a direct sum over all invertible group elements.
 It consists of:
\begin{itemize}
    \item A group \( G \) of invertible simple objects forming a pointed fusion subcategory.
    \item A single non-invertible object \( X \), whose presence extends the fusion rules beyond ordinary group multiplication.
\end{itemize}
The defining fusion relations are
\begin{align*}
X \otimes X &= \bigoplus_{g \in G} g, \\
g \otimes X &= X \otimes g = X, \qquad \forall\, g \in G,
\end{align*}
while the invertible objects fuse according to the group law of \(G\).

\noindent For \( G = \mathbb{Z}_7 = \{1, g, g^2, g^3, g^4, g^5, g^6\} \), the fusion algebra takes the explicit form
\begin{align*}
g^i \otimes g^j &= g^{\,i+j}, \qquad \text{with } g^7 = 1,\\
g^i \otimes X &= X \otimes g^i = X,\\
X \otimes X &= 1 + g + g^2 + g^3 + g^4 + g^5 + g^6.
\end{align*}
The corresponding quantum dimensions are
\[
\dim(g^i) = 1 \quad (i=1,\dots,6), \quad 
\dim(X) = \sqrt{|G|} = \sqrt{7}.
\]
\noindent The appearance of the non-invertible object \(X\) introduces a genuinely non-invertible structure into the fusion algebra, as its self-fusion produces a direct sum over all invertible sectors, a hallmark feature of Tambara--Yamagami categories.

\section{\label{app:B}Phase Rephasing of the Neutrino Mass Matrix}

\noindent Unphysical phases in a general complex symmetric Majorana neutrino mass
matrix can be removed by suitable rephasings of the lepton fields. We define
the rephasing parameters as
\begin{equation}
\alpha_i = -\arg(r_i),
\qquad i=1,2,3,
\end{equation}
corresponding to the transformation
$M_\nu \to P\,M_\nu\,P$ with
$P=\mathrm{diag}(e^{i\alpha_1},e^{i\alpha_2},e^{i\alpha_3})$.
This choice renders all $r_i$ real and reduces the
number of independent parameters.

\noindent After this rephasing, the neutrino mass matrix is given by
\begin{equation}
\resizebox{1\columnwidth}{!}{$
\begin{pmatrix}
0.1615 + 0.09065\, i & 0.2310 + 0.3808\, i & 0.1508 + 0.1480\, i \\
0.2310 + 0.3808\, i & -0.7870 + 1.598\, i & -1.130 + 0.6531\, i \\
0.1508 + 0.1480\, i & -1.130 + 0.6531\, i & -1.481 - 0.3110\, i
\end{pmatrix},
$}
\end{equation}

\noindent which is symmetric, as required for Majorana neutrinos, with physical
CP-violating phases residing in the off-diagonal elements. The diagonalization of $M_\nu$ gives the lepton mixing (PMNS) matrix
$U_{\rm PMNS}$ as
\begin{equation}
\resizebox{0.6\columnwidth}{!}{$
U_{\text{PMNS}} =
\begin{pmatrix}
0.8233 & 0.5480 & 0.1482 \\
0.4517 & 0.4972 & 0.7407 \\
0.3437 & 0.6727 & 0.6553
\end{pmatrix},
$}
\end{equation}
consistent with current neutrino oscillation data.

\end{document}